\documentclass[12pt,preprint]{aastex}
\usepackage{amsmath}
\usepackage{natbib}

\begin{document}

\bibliographystyle{apj}

\title{On the Emergent Spectra of Hot Protoplanet Collision Afterglows}

\author{Eliza Miller-Ricci}

\affil{Harvard-Smithsonian Center for Astrophysics, 60 Garden St. Cambridge,
        MA 02138}

\email{emillerricci@cfa.harvard.edu}

\author{Michael R. Meyer}

\affil{Institute for Astronomy ETH, Physics Department, HIT J 22.4, CH-8093 
       Zurich, Switzerland}

\author{Sara Seager}

\affil{Department of Earth, Atmospheric, and Planetary Sciences, Department
       of Physics, Massachusetts Institute of Technology, 54-1626, 77
       Massachusetts Ave., Cambridge, MA 02139}

\author{Linda Elkins-Tanton}

\affil{Department of Earth, Atmospheric, and Planetary Sciences, Massachusetts 
       Institute of Technology, 54-824, 77 Massachusetts Ave., Cambridge, MA
       02139}

\begin{abstract}

We explore the appearance of terrestrial planets in formation by studying the 
emergent spectra of hot molten protoplanets during their collisional
formation.  While such collisions are rare,  the surfaces of these bodies may 
remain hot at temperatures of 1000-3000 K for up to millions of years during 
the epoch of their formation (of duration 10-100 Myr).  These object are 
luminous enough in the thermal infrared to be observable with current and next 
generation optical/IR telescopes, provided that the 
atmosphere of the forming planet permits astronomers to observe brightness 
temperatures approaching that of the molten
surface.  Detectability of a collisional afterglow depends on properties
of the planet's atmosphere --  primarily on the mass of the 
atmosphere. A planet with a thin atmosphere is more readily detected, because 
there is little atmosphere to obscure the hot surface. Paradoxically, a more 
massive atmosphere prevents one from easily seeing the hot surface, but also 
keeps the planet hot for a longer time.  In terms of planetary mass, more 
massive planets are  also easier to detect than smaller ones because of their 
larger emitting surface areas -- up to a factor of 10 in brightness between 1 
and 10 M$_\earth$ planets.  We present preliminary 
calculations  assuming a range of protoplanet masses (1-10 $M_\earth$), 
surface pressures (1-1000 bar), and atmospheric compositions, for molten 
planets with surface temperatures ranging from 1000 to 1800 K, in order to 
explore the diversity of emergent spectra that are detectable. While current 
8- to 10-m class ground-based telescopes may detect hot protoplanets at wide 
orbital separations beyond 30 AU (if they exist), we 
will likely have to wait for next-generation extremely large telescopes or
improved diffraction suppression techniques to 
find terrestrial planets in formation within several AU of their host stars.
\end{abstract}


\section{Introduction}

Recent estimates suggest that between 10-20\% of Sun-like stars harbor gas 
giant planets with masses greater than 30\% the mass of Jupiter and with 
orbits ranging from $< 0.1$ AU to beyond 20 AU \citep{cum08}.  Based on new 
discoveries of super-Earths with masses 3-30 times the mass of the Earth
\citep[e.g.][]{lov06, udr07, may08} we anticipate that such bodies might be 
even more common.  We await
confirmation of these ideas from
enhanced radial velocity and microlensing surveys, as well as space-based
transit missions such as CoRoT \citep{bag03, bar05} and Kepler 
\citep{bor04, bas05}.
Recently, direct imaging surveys have produced the first widely-accepted
images of extrasolar gas giant planets \citep{kal08, mar08}, proving that 
direct detection is yet another viable planet detection method.  In addition 
to providing 
estimates of temperature and luminosity (and thus constraints on radius),
direct detection enables us to investigate the composition of the planetary 
atmosphere, complementary to studies of bulk composition.  Furthermore, the 
radiant energy of planets at orbital radii beyond that where insolation 
dominates the energy budget, provides estimates of the internal energy of the 
planet.  Together, these data provide a whole greater than the sum of their 
parts in constraining models of the formation, structure, and evolution of 
planets of all masses.  


What are the prospects for directly imaging terrestrial planets like Earth?  
Space-based mission concept studies of
1-2 meter class telescopes are underway that could just barely detect an Earth 
around the nearest stars at visible wavelengths \citep[e.g.][]{guy06}.  
Extremely large ground-based telescopes 
of the future likewise have the angular resolution to see planets around the 
nearest stars in the thermal infrared.  Someday, 
ambitious space missions like Darwin/TPF will 
provide images and spectra of Earth-like planets around dozens of 
carefully chosen targets that will revolutionize our understanding of whether 
planetary systems like our own (and the potential for life that such systems 
represent) are common or rare in the Milky Way.

An intriguing near-term possibility is to search for hot protoplanets during 
their epoch of formation, as originally proposed by \citet{ste94}.  
\citet{zah07} present a scenario for the early 
evolution of the Earth after the Moon-forming impact, thought to be the last of
a series of giant impacts that built the Earth from a swarm of protoplanets
\citep{ste87, ken04, ken06}.  Such impacts could impart enough energy on the 
forming 
protoplanet to render its surface molten.  Indeed, a 1500 K molten body of
one Earth radius is more than 600 times more luminous than the Earth, with a
thermal emission spectrum peaking at wavelengths beyond 2 microns.  The 
radiative lifetime of such a body in free space - if its atmosphere is
completely blown off during the collision --  is short 
\citep[$< 100,000$ yrs;][]{ste94} compared to the expected age of such objects
\citep[1-100 Myr;][]{cha01}.  However, such hot protoplanets may be observable
if the molten magma ocean phase of the young forming 
protoplanet can last long enough.  A long-lived magma ocean requires an 
atmosphere to slow down cooling -- possible if the protoplanet retains a 
primordial atmosphere \citep{gen03} or can release a volatile atmosphere 
through outgassing \citep{elk08}.  With an atmosphere, the planetary surface 
could remain molten for durations of up to millions of years.  

Whether or not a substantial fraction of any pre-existing atmosphere remains 
after a large collision is debated \citep{oke77, gen03}.  However, recent work 
by \citet{elk03} and \citet{elk08} describes a second longer-lasting magma
ocean phase that is naturally accompanied by a thick atmosphere.  
It is this phase
that dominates the timescale for how long the planet remains hot.  The first 
magma ocean phase results from the energy imparted during the giant impact
process. As the planet cools, the magma ocean solidifies, and volatile 
elements are released to form a new planetary atmosphere on timescales on 
the order of 100,000 years.  This solidification process is 
by its nature gravitationally unstable.  The magma ocean solidifies from the
bottom upwards, and because the lighter element Mg is preferentially 
included in solid-forming minerals over the heavier element Fe, the 
solidifying magma ocean is less dense at the bottom and more dense at the top.
This gravitational instability results in a fast overturning of the mantle, 
provided that the planet mass is larger than a lunar mass, and
the magma ocean contains more than about 4 mass percent iron, as likely 
occurred in solar system terrestrial planets.    
As the hot, low-density material begins to rise, it experiences adiabatic
decompression causing the mantle to remelt -- once again forming a magma ocean.
This is the second magma ocean phase, and this time the freshly molten planet 
is able to retain its high temperature over much longer timescales,
due to the insulating effects of the outgassed atmosphere.  This second
molten phase can persist for several to ten million years depending on the 
planet mass and composition of the outgassed atmosphere \citep{elk08}.  
 
If, as we expect, forming terrestrial planets are built through 
giant impact collisions, several post-impact magma ocean phases per 
protoplanet are likely.  And if each planetary system produces multiple 
terrestrial or super-Earth systems (such as 
HD 40307, HD 69830, or the Solar System), we might hope to catch an 
Earth-like planet in a molten phase.  A typical system with 2 super-Earths, 
each experiencing 2 giant impacts during formation that result in magma
ocean phases lasting 2 
million years each, will potentially be observable for 8 million years out of 
the 10-100 Myr years thought
required to build terrestrial planets \citep{obr07}  -- consistent with the 
timing of the Moon-forming impact from recent Hf-W chronologies 
\citep[][see also Jacobsen et al.~submitted]{kle08}.  This calculation 
results in a 10\% chance
to observe such an event if the estimates of the frequency and duration of 
events are accurate.  Indeed, \citet{mam07} explore the hypothesis that the 
enigmatic low-luminosity companion to 2MASS 1207a is in fact a hot protoplanet
collision afterglow.  In that case, the (improbable) hypothesis explored
is that of a collision between a 7 M$_\earth$ projectile and a 70 M$_\earth$
target at 55 AU orbiting a 5-10 Myr brown dwarf found in a young cluster.  

The presence of a hot molten surface however is a necessary but not 
sufficient condition for detection.  Atmospheric windows must be present in the
planet's emitted spectrum 
that enable astronomers to observe brightness temperatures approaching 
that of the planetary 
surface.  Addressing this issue is the subject of the present work.  We 
present calculations of the emergent spectrum of a hot protoplanet 
with a molten surface, exploring a range of planet masses, atmospheric mass 
fractions, atmospheric compositions, and surface temperatures.  We pay 
particular attention to the ground-based 
astronomical windows in the near-infrared where current and future 
telescopes capable of detecting these hot protoplanet collision afterglows are 
expected to be particularly effective.   Our calculations suggest that, even 
in the most favorable assumptions, hot young planets are likely not detectable 
by current instruments on existing telescopes. However, instrumentation and 
space observatories 
under construction should be able to detect a subset of
such objects on orbits beyond 20 AU (if they exist in sufficient numbers), 
while the next generation of extremely large telescopes will survey
orbits comparable to the terrestrial planets in our solar system.

\section{Methodology}

\subsection{Range of Planetary Parameters Considered \label{parameters}}

A handful of super-Earths in the 1-10 M$_\earth$ range have already been 
discovered, and it is expected that many more will be reported in the coming
years.  Observational constraints on super-Earth atmospheres however are not
yet available.  In the absence of a large statistical sample, we 
consider super-Earths with atmospheres covering a 
wide range of parameter space that could be expected for such planets.
In terms of planetary mass, we consider super-Earths of
1, 5, and 10 M$_\earth$.  Assuming they have a similar 
composition to that of the Earth (67.5\% silicate mantle and 32.5\% iron core),
these planets will have corresponding surface gravities, $g$, of approximately 
9.8, 21.8, and 29.6 m/s$^2$ according to theoretical mass-radius relationships 
for solid exoplanets \citep{sea07}.  These values for $g$
depend on the assumed composition of the solid portion of the planet
but only vary by up to 40\% if vastly different compositions are employed such 
as pure water or pure iron.  

The atmospheric composition  of a super-Earth depends strongly on the 
conditions leading to its creation.  Factors such as accretion and
outgassing history, which are difficult to constrain with models, will 
ultimately determine the composition \citep{elk08b}.  Additionally, molecular 
abundances in
the atmosphere can be further altered through processes such as 
photochemistry, atmospheric escape, and interactions between the atmosphere 
and the planetary surface.  For this reason we choose to examine the spectral 
signatures of a range of atmospheres that span the possible outcomes from 
ougassing and accretion scenarios.  The atmospheres we consider are: 

\begin{enumerate}

\item \textit{Solar Composition Atmosphere} -- This is our benchmark 
case, which represents either the remains of an initial accreted
atmosphere or a hydrogen-rich outgassed atmosphere.  We employ solar 
elemental abundances \citep{asp05} and assume that the 
molecular constituents reside in a state of chemical equilibrium (see
Section~\ref{model}).  The resulting atmosphere is composed by volume
of 85\% H$_2$, 15\% He, and 560 ppm H$_2$O.  At the temperatures that we
are considering, this atmosphere bears strong similarities to a brown dwarf, 
although the surface gravity here is far lower.
\item \textit{30}$\times$ \textit{Solar Enhanced Metallicity Atmosphere} --
This case is similar to scenario 1, but here the abundances of
all elements other than hydrogen and helium have been enhanced to 30 times 
that of the Sun resulting in an atmosphere that is 81\% H$_2$, 15\% He, and 
1.7\% H$_2$O. 
\item \textit{90\% Water (steam), 10\% CO$_2$ Atmosphere} -- This atmosphere, 
as well 
as the next two that follow, represents the possible compositional outcome of 
an outgassing scenario.  H$_2$O-CO$_2$ atmospheres are expected to result
naturally from surface outgassing as the planet cools.  However, the exact 
ratio of H$_2$O to CO$_2$ depends strongly on the initial volatile content
of the hot protoplanet \citep{elk08b}.  The three outgassing scenarios on this 
list -- scenarios 3, 4, and 5 -- encompass
a probable range of H$_2$O-CO$_2$ atmospheres that may occur.
\item \textit{50\% Water (steam), 50\% CO$_2$ Atmosphere}
\item \textit{10\% Water (steam), 90\% CO$_2$ Atmosphere}
\item \textit{Venus-Composition Atmosphere} -- This atmosphere has a 
composition mirroring that of Venus' atmosphere and could be obtained either by
outgassing or by accretion with subsequent escape of light-weight elements.
Here the atmosphere is composed predominantly of CO$_2$ -- 96.5\% by volume -- 
along with 3.5\% N$_2$ and 20 ppm H$_2$O \citep{lew95}.


\end{enumerate}

Protoplanet collision afterglows will only be observable in cases where the
heat flow from the planet's interior is sufficient to 
sustain a high emergent flux from the top of the planetary atmosphere.  This 
is likely to occur only when the mantle is in a low-viscosity state, requiring 
a full to partial melt of the mantle.  For this reason, we consider surface
temperatures that are indicative of a molten planetary surface.  Immediately 
after overturn the surface temperature of a (1 M$_\earth$) terrestrial planet 
will vary from $\sim$ 3000 - 1500 K, and will slowly cool  from there, becoming
solid around 1000 K \citep{elk08}. The more massive the planet, the hotter the 
solidus temperatures become at depth.  There is some evidence that at depth 
silicate liquids become denser than their coexisting solids \citep{sti05}, a 
condition that
would mark the bottom of the magma ocean that would later interact with the 
surface.  This may limit the temperature of overturning mantle materials to 
about 4000 K, even for the most massive super-Earths.  Following the mantle 
overturn, the surface cools over millions of years through 1000 K, at which 
temperature an Earth-like composition would be solid.  We produce emergent 
spectra from the intermediate to low end of 
the molten temperature range, where the planets will spend most of their 
time as they cool.  The four surface temperatures that we consider are 
1800, 1500, 1200, and 1000 K.  It serves pointing out that we are considering 
only rocky terrestrial planets here and not ice planets, which would form 
outside of the snow line and then potentially migrate inwards.  These planets 
would also experience similar protoplanet collisions, and whether or not they
retain high enough brightness temperatures to be observable will depend 
strongly on how much energy is imparted in the collision and how quickly this 
energy is reradiated to space.

The surface pressure conditions for protoplanet collision afterglows are 
difficult to ascertain, since the pressure is dependent on atmospheric mass, 
which is not known
\textit{a priori} and is strongly model dependent.  Outgassed atmospheres of 
super-Earths may be able to attain very high 
surface pressures of up to several thousand bars \citep{elk08b}, depending on 
their exact histories and the volatile content upon formation.
We therefore produce spectra for atmospheres covering a range of probable 
surface pressures -- 1, 10, 100, and 1000 bars.

\subsection{Description of the Model Atmosphere \label{model}}

To determine the emitted protoplanet spectra we place a simple model
atmosphere on top of the hot molten planetary surface.  Our atmosphere model
follows the one developed in \citet{mil08} but differs in that here the
atmosphere is almost entirely heated from below by the planet's hot surface
rather than by the irradiation from the host star.     
The atmosphere that we consider is a grey gas in hydrostatic equilibrium.
The temperature structure is then given by
\begin{equation}
T^{4} = \frac{3}{4} T_{eff}^{4} (\tau + \frac{2}{3}),
\label{grey}
\end{equation}
where $\tau$ is the optical depth, and $T_{eff}$ is the planet's effective 
temperature.  The pressure structure is determined by integrating the
equation of hydrostatic equilibrium:
\begin{equation}
\frac{\rm{d}P}{\rm{d}\tau} = \frac{g}{\kappa_{gr}},
\label{hydro}
\end{equation}
where $g$ is the surface gravity and $\kappa_{gr}$ is the mean opacity in 
units of cm$^2$/g, based on the chemistry that we have employed.  We calculate 
Rosseland mean opacities at depth but switch
to Planck mean opacities for $\tau \ll 1$.  Convection is included in the lower
atmosphere, by switching to an adiabatic temperature-pressure profile in 
regions that are convectively unstable.  For our entire calculation we assume 
an ideal gas equation of state (EOS), despite the fact that water is known to
deviate from an ideal gas -- particularly at high pressures and 
abundances.  For the hot protoplanet we 
specify our desired surface temperature and pressure as boundary conditions, 
and then adjust $T_{eff}$ and the upper limit on the optical depth scale 
($\tau_{surf}$) until they agree with those conditions.  The other free
parameters, $g$ and $\kappa_{gr}$ are specified by the model assumptions -- 
the mass and radius of the planet and its atmospheric composition (see
Section~\ref{parameters}, above).  

\begin{figure}
\plotone{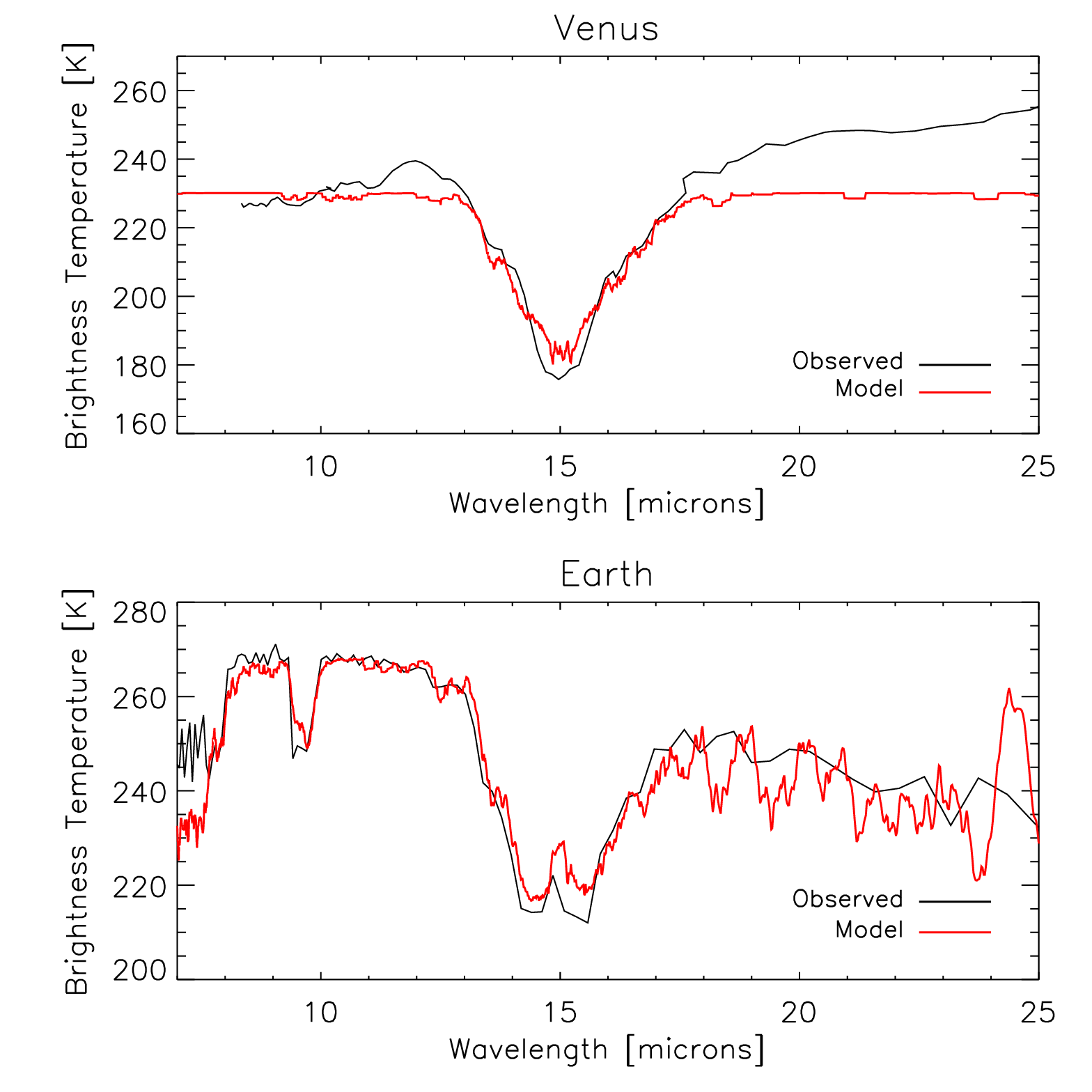}
\caption{Tests of our model scheme to reproduce Venus' \citep{mor86} and 
        Earth's \citep{chr97} emitted
        spectra.  To simulate Venus' sulfuric acid cloud layer, we cut off
	all emission below an altitude of 70 km ($T_B$ = 230 K).  This 
	explains the mismatch between the model and the 
	observations in the continuum at wavelengths longer than 18 $\mu$m,
	where the brightness temperature becomes as high as 255 K.  For Earth
	we simulate the planet-averaged emission spectrum by averaging together
	a cloud-free model spectrum with a cloudy spectrum (emission cut off
	at a cloud deck of altitude 6 km), assuming that 59\% of Earth's 
	surface is obscured by clouds.  For both planets, our
	model reproduces the observations to within 10\%, even in the continuum
	of Venus' spectrum.
        \label{solarsystem}}
\end{figure}

After determining the temperature-pressure profile, we calculate the planet's 
emitted spectrum -- at a resolution of 1,000 -- by integrating the equation of 
radiative transfer through the 
planet's atmosphere.  The emergent intensity, $I$ is given by
\begin{equation}
I(\lambda,\mu) = \frac{1}{\mu}\int^{\tau}_{0} S(T) e^{-\tau'/\mu} d\tau',
\end{equation}
where $S$ is the source function (assumed here to be Planckian) , $\mu$ is the 
cosine of the viewing angle, and $\tau$ is the optical depth at the base of 
the atmosphere.  We have tested this scheme to reproduce Earth's and Venus' 
emitted spectra, given their known T-P profiles and simulating Venus' opaque 
H$_2$SO$_4$ cloud deck by cutting off the planetary emission at an altitude of 
70 km above the surface.  Our resulting spectra are in agreement with the 
observed planet-averaged emission for both of these cases \citep{mor86, chr97}
to within a factor of several to ten percent -- sufficient for modeling 
the observational signals of extrasolar planet atmospheres -- although
difficulties remain for models to exactly reproduce Venus's IR spectrum
\citep[e.g.][]{pol93} (see Figure~\ref{solarsystem}).

In computing the emergent spectra, we include the dominant sources of 
molecular line opacity from 0.1 to 100 
$\mu$m in the IR -- CH$_4$, CO \citep[][and references therein]{fre07}, 
CO$_2$ \citep{rot05}, and H$_2$O \citep{fre07, par97}.  For the line profiles,
we employ a Voigt broadening scheme.  In addition to the line opacities,
for a Venus-like atmospheric composition
(case 6 above) CO$_2$ - CO$_2$ collision induced opacities become important, 
and without them the CO$_2$ line opacities imply the presence of wide 
transparent windows in the near-IR.  Unfortunately, a full prescription for
CO$_2$ - CO$_2$ collision induced opacities is not available in the literature,
and we instead interpolate between near-IR laboratory measurements at 2.3 
$\mu$m \citep{bro91} and theoretical calculations for wavelengths longward
of 40 $\mu$m \citep{gru97}.  

For the solar and enhanced metallicity atmospheres (cases 1 and 2) we
determine molecular abundances in chemical equilibrium, starting from the 
initial elemental abundances for these two cases.  Chemical equilibrium should
be a reasonable assumption here given the high temperatures in the lower 
atmosphere, implying that reactions will occur quickly and will not be limited
by temperature.  However, in the absence of effective mixing, photochemical
processes may drive the composition away from equilibrium in the upper 
atmosphere.  For the equilibrium calculation, we perform a
minimization of Gibbs free energy for 172 gas phase molecules and 23 atomic 
species, following the method outlined in \citet{whi58}.  
The Gibbs free energy of each molecule as a function of temperature
is paramaterized by a polynomial fit from \citet{sha90}, which we have found to
be well-matched to data from the NIST JANAF thermochemical tables 
\citep{cha98} over our temperature range of interest of 200-1500 K.  We do not
include condensed species in our equilibrium chemistry model, however for each 
atmosphere scenario we do check whether the temperature-pressure profile for
each major species crosses into the regime where condensation and cloud 
formation would be expected.

\section{Hot Protoplanet Emission Spectra}

\subsection{Main Results \label{results}}

The emergent spectra for hot protoplanets with 1500 K $T_{surf}$ are shown in
Figure~\ref{spectra} (in terms of brightness temperature) and Figure~\ref{flux}
(in terms of flux for a planetary system at a distance of 30 pc).  
There are spectral windows where the modeled fluxes are quite high, and the
potential exists to observe brightness temperatures up to the planet's 
surface temperature.  However, this is generally only the case for the 
thinnest atmospheres and even then only in fairly narrow atmospheric windows.  

For more massive atmospheres, the 
planet's brightness temperature drops off quickly due to increased 
surface obscuration by a larger optically thick atmosphere.  For atmospheres
with kilobar surface pressures, the maximum brightness temperature that can 
potentially be observed for any of the atmospheric compositions we have 
considered is approximately 1000 K for planets with 1800 K surface temperatures
and only 500 K for planets with 1000 K T$_{surf}$.  Unfortunately, planets with
thin atmospheres are also expected to retain their molten surfaces for the 
shortest periods of time.  More massive
atmospheres serve to prevent the escape of heat from the planetary surface
and can therefore maintain high surface temperatures for up to several million 
years \citep{elk08}.  
This is a dilemma, as the atmospheres that are the easiest to
detect owing to their high brightness temperatures will potentially be the 
least likely to be observed due to their short lifetimes.  

The effect of planetary mass on the emitted spectra is more subtle than that
of surface pressure.  According to Figure~\ref{spectra}, the 10 M$_\earth$ 
planets emit at slightly higher brightness temperatures than the corresponding 
1 M$_\earth$ 
planets -- at all wavelengths.  This effect ultimately results from the 
higher surface gravity and correspondingly smaller pressure scale height 
on the more massive super-Earth.  For a given surface pressure, as the mass
of the planet increases, the thickness of the atmosphere that the observer
must ``look through'' to see the planetary surface drops off -- explaining 
the somewhat higher brightness temperatures for more massive
planets.  However, for a given mass of atmosphere, the surface pressure will
be higher on more massive planets owing to their correspondingly higher surface
gravities.  In terms of emergent flux, the more massive super-Earths are 
clearly more readily observable (Figure~\ref{flux}) because they emit from a 
larger surface area.  The emitted fluxes are therefore weakly dependent on the 
composition of the planet itself, owing to
its effect on the planetary radius.  Here we have employed values of 1, 1.5,
and 1.8 R$_\earth$ for the 1, 5, and 10 M$_\earth$ planets, respectively,
corresponding to a planetary composition similar to that of the Earth. 

Figure~\ref{temp} shows the dependence of the emergent spectrum on the planet's
surface temperature for the case of a 1-bar solar-composition atmosphere.  As
expected, the brightness temperature is strongly dependent on the surface
temperature, with hotter planets emitting higher fluxes.  Generally, for all of
the atmospheres we have studied, planets with surface temperatures of 1500, 
1200, and 1000 K emit at brightness temperatures that are up to 20\%, 40\%, 
and 50\% reduced relative to that of an 1800 K $T_{surf}$ planet.  This is in
line with an expectation that the planet's brightness temperature should scale 
linearly with surface temperature, as a direct consequence of 
Equation~\ref{grey}

\subsection{Description of the Spectra}

\subsubsection{Cases 1 and 2 -- Solar Composition and Enhanced Metallicity}

For the solar composition and
enhanced metallicity atmospheres (scenarios 1 and 2 from 
Section~\ref{parameters}), spectral features mostly result from water
absorption bands.  Additionally, in chemical 
equilibrium, the main carbon-bearing species expected for these atmospheres 
is methane, which reveals itself through absorption features in the spectra at 
both 2.2 and 3.4 microns.  As an atmospheric constituent, methane is 
particularly susceptible to UV photodissociation.  Depending on 
the planet's proximity to the host star and the properties of the stellar UV 
emission, methane may or may not be stable in the hot super-Earth 
atmosphere.  Additionally, the presence or absence of methane depends strongly
on the ratio of its photodissociation rate to the rate of return 
reactions in the planetary atmosphere, which is strongly temperature dependent.
The spectral features at 2.2 and 3.4 microns can therefore be used to diagnose 
whether or
not methane is present in a hot super-Earth atmosphere, but it would not
be surprising to find this molecule out of equilibrium.  

If methane is successfully destroyed, the resulting carbon atoms will reform 
into the more photochemically stable molecules CO and CO$_2$.   Even in a 
state of chemical equilibrium, the enhanced metallicity atmosphere may contain
enough CO$_2$ for its spectral fingerprint to be observable at 4.3 $\mu$m, as 
is the case for the 1-bar atmosphere -- see Figure~\ref{spectra}.  While CO is
expected to be present in these atmospheres at abundances 100 times greater 
than that of CO$_2$ (up to 1ppm), it does not display
any strong absorption bands shortward of 4.5 $\mu$m in these models.

\subsubsection{Cases 3, 4, and 5 -- H$_2$O - CO$_2$ Atmospheres}

The spectra for the three outgassed atmosphere scenarios --
3, 4, and 5 -- all strongly resemble one another owing to saturation of the
water vapor and CO$_2$ lines at lower partial pressures than the ones
considered here.  This is despite the fact
that the CO$_2$ abundance increases from 10\% to 90\% and the H$_2$O abundance 
correspondingly decreases from 90\% to 10\% across these three cases.  
For cases 3, 4, and 5, the spectra reveal absorption features of both
water and CO$_2$.  Absorption bands at 1.1, 1.4, and 1.9
$\mu$m all result from water, while the band at 4.3 $\mu$m is due to 
CO$_2$.  The additional absorption feature around 2.7 $\mu$m results from
overlapping H$_2$O and CO$_2$ bands.  Due to the strong spectral features 
at such high
abundances of water and CO$_2$, an atmosphere composed of these two molecules
should be readily identifiable.  However, discerning the exact abundances of
H$_2$O and CO$_2$ from emission spectra is more difficult, since the changes
in the features are those realized in the non-linear wings of the absorption
profiles.  

\subsubsection{Case 6 -- Venus Composition}

Hot super-Earth atmospheres of Venus-like 
composition (scenario 6 from Section~\ref{parameters}) achieve high brightness
temperatures, approaching that of the planetary surface, across large portions
of the IR spectrum.  For this reason, Venus-composition atmospheres probably 
have the best chance of being observed from the ground, although escape of
flux through the transparent atmospheric windows will result in these planets
cooling more quickly than some of the other composition scenarios.  Since 
these atmospheres are composed predominantly of CO$_2$, their molten 
surfaces are not obscured by strong water absorption bands in the near-IR.  
The primary spectral features for a Venus-like atmosphere result from CO$_2$,
but there are large windows between absorption bands from 2 to 2.5 microns
and from 3 to 4 microns where an observer could potentially detect emission
from the planetary surface.  In between the CO$_2$ absorption bands the main
source of opacity is collisional.  As mentioned above in Section~\ref{model},
a full characterization of CO$_2$ - CO$_2$ collision induced opacities is
not available in the literature, and we have therefore interpolated between
the only two available datasets.  Artifacts as a result of this interpolation 
can be seen in the bottom right hand panel of Figure~\ref{spectra} with the 
flat ``ceiling'' in all of the spectra.  
It is possible that the true values for the CO$_2$-CO$_2$ collision 
induced opacities in this wavelength range could be quite different from the 
ones we have used for this study.  However, additional laboratory 
experiments will clearly be necessary to resolve this issue.

\subsection{The Effect of Clouds}

The spectra presented in Figures~\ref{spectra} and~\ref{flux} are all for 
cloud-free atmospheres.  However, the presence of clouds can further affect 
the observability of a molten super-Earth by preventing the escape of flux from
the planet's hot surface.  Generally, if clouds are present, the
atmosphere will emit radiation at or below the
temperature of the cloud deck.  Venus is an good example of this effect, 
where the planetary surface is very hot, but the IR emission from the planet
occurs at a much lower temperature owing to the fact that most of the flux
emitted at the surface does not escape through Venus' sulfuric acid cloud 
layer. 

For the atmospheric scenarios that we present here, the temperatures are too
high for CO$_2$ clouds to form.  However, water clouds are a possibility
in a number of cases.  One-dimensional models like the one we present here
have a limited ability to properly account for the presence of clouds.  For
example, a 1-D model would predict water clouds on Earth, but it
would not correctly predict that these clouds are patchy and do not 
entirely obscure the surface of the Earth at any given time.  In our model
we determine if water clouds are present by comparing the temperature-partial
pressure curve for H$_2$O against its condensation curve.  If the two curves
intersect, we assume that clouds can form at the lowest location in the 
atmosphere where water will condense.  

We find that most of the atmosphere
scenarios with high abundances of water have the potential to form clouds.
For atmospheric composition scenarios 1, 2, 3, 4, and 5 (from 
Section~\ref{parameters}), clouds will form for 100- and 1000-bar
atmospheres (as well as for 10-bar atmospheres in scenarios 2, 3, and 4) at 
temperatures below 320 K and pressures below 0.1 bars.  If 
these clouds are thick and entirely blanket the planetary surface, then
the observability of these planets will be severely limited due to their 
resulting low brightness temperatures.  However, if the clouds only obscure a
fraction of the planet, quite high planet-averaged brightness temperatures 
can still be achieved.  This is shown in Figure~\ref{clouds} 
where we plot full- and partial-cloud cover spectra for all of the atmospheres 
where clouds are a 
possibility.  The presence of clouds does have an interesting additional
consequence in that clouds can essentially insulate the planet, allowing the
planetary surface to remain molten for much longer than it would be able
to otherwise.

\section{Observational Implications}
\label{sec-implications}

\subsection{Prospects With Current Telescopes}

Given these results, we now explore the prospects of detecting hot 
protoplanet collision afterglows with current instrumentation and offer ideas 
concerning future search strategies.  Several groups
\citep[e.g.][]{laf07, bil07, kas07b, hei08} have published null results for
solar-type stars based 
on high contrast imaging results in the near-infrared at 1.65 microns (H-band) 
and 3.6 microns (L-band).  The goal of these surveys was to detect gas giant 
planets at large separations and place limits on the power-law surface density 
of such companions, as well as on the outer limit of circumstellar planet
formation, based on extrapolation of results from radial velocity surveys 
\citep[e.g.][]{cum08}.  Results to date suggest that massive gas giant planets 
($>$ 3 M$_J$) do not form frequently at separations beyond 20 AU, compared to 
gas giants at smaller separations.  The question then arises - could such 
surveys have detected the hot 
protoplanet collision afterglows, given the model spectra described above?

Forming super-Earths with extremely 
tenuous atmospheres (the 1 bar models of Figure~\ref{spectra}) would be 
detectable if they lurk at large enough angular separations.  For example, the 
H-band observations of \citet{laf07} from the Gemini Deep Planet Survey are 
limited by the contrast of any 
potential planet against the glare of the central Sun-like star at separations
within several arcseconds.  In order to detect a 2 R$_\earth$ hot protoplanet 
in formation, with a solar composition and a 1 bar atmosphere, a 
contrast of 14.3 magnitudes is required in the H-band.  \citet{laf07} achieve 
this contrast at a typical angular separation of 1.5 arcseconds. The youngest
targets in their sample (with ages less than 100 Myr) have typical distances 
of 30 pc.  As a result, these observations are sensitive to hot protoplanet 
collision afterglows on physical scales greater than 45 AU in these systems.

For comparison, the L-band observations taken at the VLT as reported by 
\citet{kas07b} reach the 
thermal background limit (rather than being contrast-limited) at angular 
separations greater than 1.5 arcseconds.  This corresponds roughly to 10 times 
the classical Rayleigh diffraction-limit.  The background limit reached in 
15-20 minutes of on-source integration time was L $<$ 16 magnitude.  Based on 
the models presented above, a 2 R$_\earth$ hot protoplanet collision 
afterglow would have an absolute L-band magnitude of 15.5 corresponding
to a distance of 10 pc while a forming 1 R$_\earth$ planet would have 
M$_L$ = 17 magnitudes.  As the targets in their survey range in age from 
10-30 Myr with distances from 10-50 pc they could have in principle detected 
protoplanets in formation beyond 15 AU for their nearest targets or hot
planets as large as Uranus (although such planets would probably
not be primarily composed of molten rock) out to distances of 30 pc
at orbital distances beyond 45 AU.

No extremely faint common proper motion companions have been detected in any 
of these surveys, perhaps indicating that in addition to gas giant planets, 
the formation of super-Earths at large distances from their
host stars may be rare as well.  This would not be surprising, considering
that planet formation through giant impact accretion is expected to be 
inefficient at the large orbital separations probed by current surveys.  
Additionally, as mentioned above in Section~\ref{results}, \citet{elk08} points
out that there is a relationship between the length of time a forming planet's 
surface remains molten and the density of its insulating atmospheric blanket.  
Those protoplanets with thin atmospheres (1-100 bars) retain surface 
temperatures above 1000 K only a little longer than the thermal cooling
time (approximately 10,000 yrs).  Protoplants with dense atmospheres will 
remain molten for up to 100 times longer.  Yet it is precisely these planets 
that are difficult to detect as the observable brightness temperatures are 
factors of 2-4 cooler than the surface temperatures (see Figure~\ref{spectra}).
For example, even a 10-bar atmosphere, which does not appreciably affect the 
cooling time, would be about 1.5 magnitudes fainter for both a 1 and 10 
M$_\earth$ planet with a solar composition atmosphere.  The difference is
even worse for the 30$\times$ solar composition atmosphere, but
not as bad for the various H$_2$O and CO$_2$ mixtures considered above.  Venus 
composition is most favorable in this regard -- even a 1000-bar atmosphere is 
only 3.7 magnitudes fainter in both the H- and L-bands.  However a 
Venus-composition atmosphere will cool more quickly than the other 
compositional scenarios since it lacks the greenhouse trap of water vapor.

Given the ratio of cooling time to age, we would need a
sample of thousands of stars to see just one protoplanet collision afterglow 
with current instrumentation.  Even then, successful detection would rely on 
the presence of collisional afterglows at large orbital separations 
($>$ 45 AU) -- an implausible scenario.  However, prospects for detections are 
far better with future instrumentation, as described below.
 
\subsection{Prospects With Future Instrumentation}

\citet{gra07} describe the capabilities of Gemini Planet Imager (GPI) under 
development for the 8 meter Gemini telescopes.  They expect to achieve 
extraordinary contrasts of more than 21 magnitudes in the H-band at 0.5 
arcsecond separations.  This would permit detection of forming terrestrial 
planets of solar composition with ease at physical separations greater than 
5-15 AU around the nearest targets.  It could even probe systems with 
relatively dense atmospheres, greatly increasing the chance for detection as 
discussed above.  Similar capabilities are planned for the VLT utilizing the 
Spectro-Polarimetric High-contrast Exoplanet REsearch (SPHERE) instrument 
\citep{beu08}.  In the L-band, \citet{ken07b} describe the 
use of an apodizing phase plate to achieve diffraction suppression, enabling 
one to reach the background limit at 3 $\lambda / D$ rather than 10 
$\lambda / D$.  This would improve the inner working angle and thus the 
physical resolution of existing surveys using current 6-10 meter telescopes by 
a factor of three.  Future work with the Large Binocular 
Telescope \citep{hin08} would enable these techniques to be used with the 23 
meter baseline of the Fizeau interferometer.  This could provide physical 
resolution within a few AU for the nearest targets using the LMIRCam
instrument now under construction \citep{wil08}.

\begin{figure}
\plotone{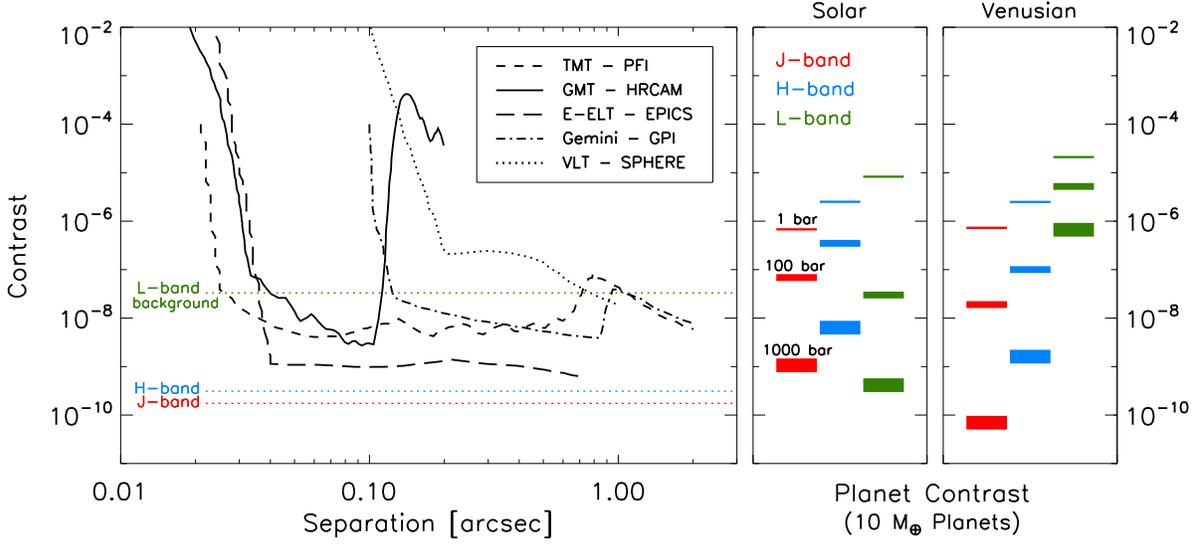}
\caption{Left - Direct imaging contrast vs.~angular separation for next 
        generation exoplanet imaging instruments.  Contrasts corresponding to 
	the sky background for one-hour integrations for the GMT on a 
	Sun-like star at 10 pc is
	overplotted in color (dotted lines) for J, H, and L band, showing that
	observations will be background limited at longer wavelengths.  
	The sensitivity curves for the TMT's planet imager PFI and Gemini's 
	GPI are from \citep{mac06} for a 4th magnitude target in H-band.  The 
	curve for 1-hour exposures of HRCAM on the GMT is also for H-band (GMT 
	Conceptual Design Report: http://www.gmto.org/CoDRpublic).  
	Corresponding sensitivity curves for SPHERE on VLT and EPICS on E-ELT 
	\citep{kas08} are instead shown at J-band.
	Right - Upper limits in contrast needed to detect hot protoplanet 
	afterglows with various atmospheric masses.  The upper limits shown 
	here are for the detection of 10 M$_\earth$ planets 
	($T_{surf}$ = 1500 K) with solar- and 
	Venus-composition atmospheres (composition cases 1 and 6) orbiting a 
	solar-type star.  The line weights denote the planet's atmospheric 
	surface pressure - 1 bar (thin lines), 100 bar (medium-weight lines), 
	and 1000 bar (thick lines).  Limits for 10 bar atmospheres have been 
	omitted to avoid confusion.  Planet-star contrast levels for a 1 
	M$_\earth$ planet are a factor of $\sim 3-10$ lower than what is shown 
	here for a 10 M$_\earth$ planet. 
	\label{large_telescope}}
\end{figure}

Further progress will be made with next-generation ground-based observatories.
One of the main science goals of future extremely large telescopes (ELTs) with 
planned mirror sizes of 25-50 m is to directly image extrasolar planets, 
perhaps even terrestrial planets.  With the increased angular resolution 
resulting from their large diameters, planet imaging instruments on telescopes 
such as the Giant Magellan Telescope (GMT), the Thirty Meter Telescope (TMT), 
and the European ELT (E-ELT) will probe probe even smaller angular separations 
than is possible with current observing facilities.  Near-infrared imagers on 
these telescopes, fed by extreme adaptive optics, would have the capability to 
detect planets within 0.1'' of potential planet-hosting stars\footnote{GMT 
Conceptual Design Report: http://www.gmto.org/CoDRpublic} \citep{mac06, kas08}.
This will allow astronomers to directly image planets within several AU of 
their host stars for the first time -- a region where we know terrestrial 
planet formation was effective in our own solar system.  Achievable contrasts 
for instruments
such as HRCam on GMT, the Planet Formation Imager (PFI) on TMT, and the 
Exo-Planet Imaging Camera and Spectrograph (EPICS) on E-ELT are predicted to 
reach factors of 10$^{-8}$ to 10$^{-9}$ ($> 20$ mag) within 0.1'' and remain 
contrast limited for Sun-like target stars within 50 parsecs ($<$ 5 AU) for 
reasonable 6-10 hour integrations.  This would permit detection of hot 
protoplanet atmospheres of  up to 100 bars for 1 M$_\earth$ planets and 
potentially up to 1000 bars for 10 M$_\earth$ planets or for planets with
surface temperatures exceeding 1500 K (see Figure~\ref{large_telescope}).  In 
the thermal infrared, diffraction suppression techniques currently being
developed \citep[e.g.][]{ken07b} might enable background-limited 
performance without the use of extreme adaptive optics within 3 $\lambda$ / D. 
For the GMT, this would result in a survey
depth of L $<$ 22 magnitudes (5 sigma, 1 hr) within 0.1" (1-5 AU for the 
nearest targets from 10-50 pc), and could be improved upon even further for 
longer integration times.  

ELTs should therefore be able to detect young forming super-Earths with thick 
atmospheres (of up to 1000 bars) within several AU of their host stars
throughout the near-IR ground-based windows from 1-5 microns.  
With the 100 $\times$ longer 'shelf-life' of these atmospheres (relative to
more tenuous 1 bar atmospheres), one 
could reasonably expect 10 \% of Sun-like stars with ages 10-100 Myr to have a 
planet in a molten surface phase at a given time.  Thus for future ELTs, 
a sample of 100 targets within 50 pc could yield several hot protoplanet 
collision afterglow candidates.  Multi-wavelength photometry and/or 
spectroscopy could then be used to determine both the temperature and 
luminosity of these objects, enabling estimates of their radii.  Spectroscopic 
follow-up could also provide constraints on composition and
surface gravity, further constraining their nature.

In terms of space-based observing, ten years from now JWST will be surveying  
nearby stars for planets at modest separation using a variety of direct 
imaging and coronagraphic techniques.  The three imaging instruments (NIRCam, 
FGS/TFI, and MIRI) will have unmatched sensitivity to search for cool planets 
in the thermal infrared \citep{mey07}.  As a result, JWST will be 
unsurpassed at large separations (beyond 1") limited only by the low 
background of a cooled space telescope.  With 3.6 micron sensitivity of 
approximately L $<$ 26 magnitude (10 sigma, 3 hr), NIRCam will be able to see 
1 R$_\earth$ planets after impact with 1000 bar atmospheres, but only at 
separations $>$ 20 AU for typical targets.  If super-Earths in formation at 
separations beyond 15 AU are there to be found (which remains debatable) it 
may be that JWST, 
with its ability to detect even the faintest forming protoplanets with the 
densest atmospheres (and thus longest 'shelf-life' for detection), will 
find the first one.

\section{Summary and Discussion}

We have shown that hot protoplanet collisional afterglows may be observable
from the ground with next generation ELT's, under certain
conditions described as follows.  Generally, the most massive terrestrial 
planets will have the
largest observable signals, due to their larger emitting surface areas.  
Planets of 10 $M_\earth$ can be up to a factor of 10 brighter than 1 $M_\earth$
planets in the near-IR.  In terms of their atmospheres, hot young super-Earths 
with atmospheres of 1-1000 bars and surface temperatures greater than 
1500 K will be detectable with next-generation 
ground-based facilities such as the GMT, TMT, and E-ELT.  However, true Earth 
analogues (1 $R_\earth$) planets with 1000-bar atmospheres may remain out of 
reach.  Thick atmospheres have the advantage that 
they insulate the planetary surface from heat loss, allowing a super-Earth in 
formation to remain molten for up to millions of years.  However, these 
planets will also be difficult to observe for the same reason -- the 
blanketing atmosphere keeps the planetary surface hot by preventing large 
amounts of flux from escaping to space.  Still, there is a compelling case to 
search for hot protoplanet afterglows with future telescopes and planet-finding
instruments.  If such
protoplanet collision afterglows are successfully observed, they will allow 
astronomers to study formation mechanisms and occurrence rates for planets in 
the low-mass, terrestrial regime.  As discussed above, successful detection
of several protoplanet collision afterglow candidates with next-generation 
ELTs should require observations of a sample of $\sim$100 young stars within 
50 pc.   

\begin{figure}
\plotone{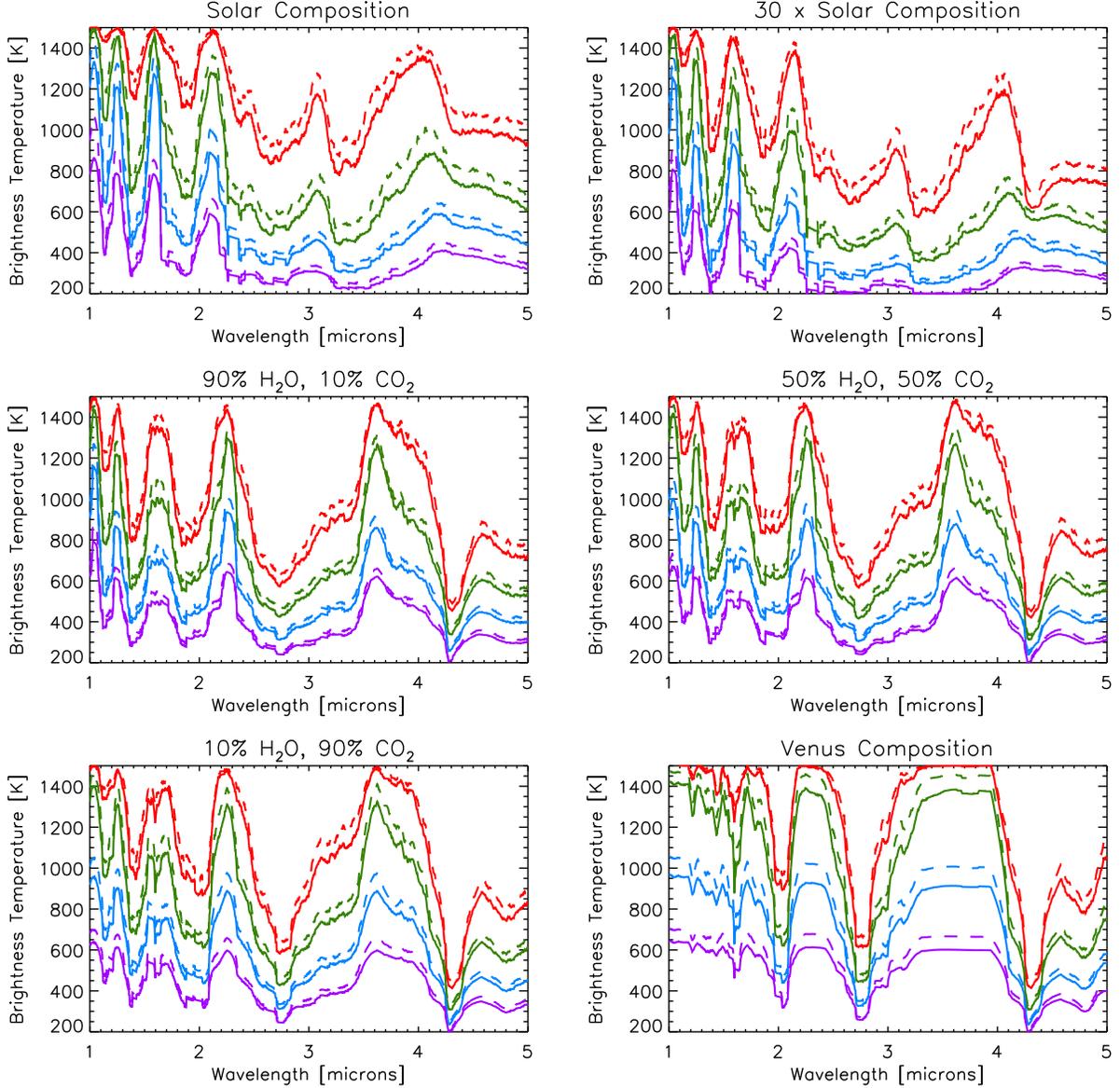}
\caption{Spectra of hot molten super-Earths for 6 cases of atmospheric
        composition.  All spectra are for planets with a 1500 K surface 
	temperature.  Planets with surface pressures of 1, 10, 100, and 1000 
	bars are denoted by red, green, blue, and purple lines respectively.  
	Spectra for 1 M$_\earth$ planets are shown with solid lines and those
	for 10 M$_\earth$ planets with dashed lines.  The spectra are 
	dependent on the underlying temperature-pressure 
	profile, and profiles other than a simple gray approximation may 
	therefore alter the appearance of the spectra presented here.
        \label{spectra}}
\end{figure}
\begin{figure}
\plotone{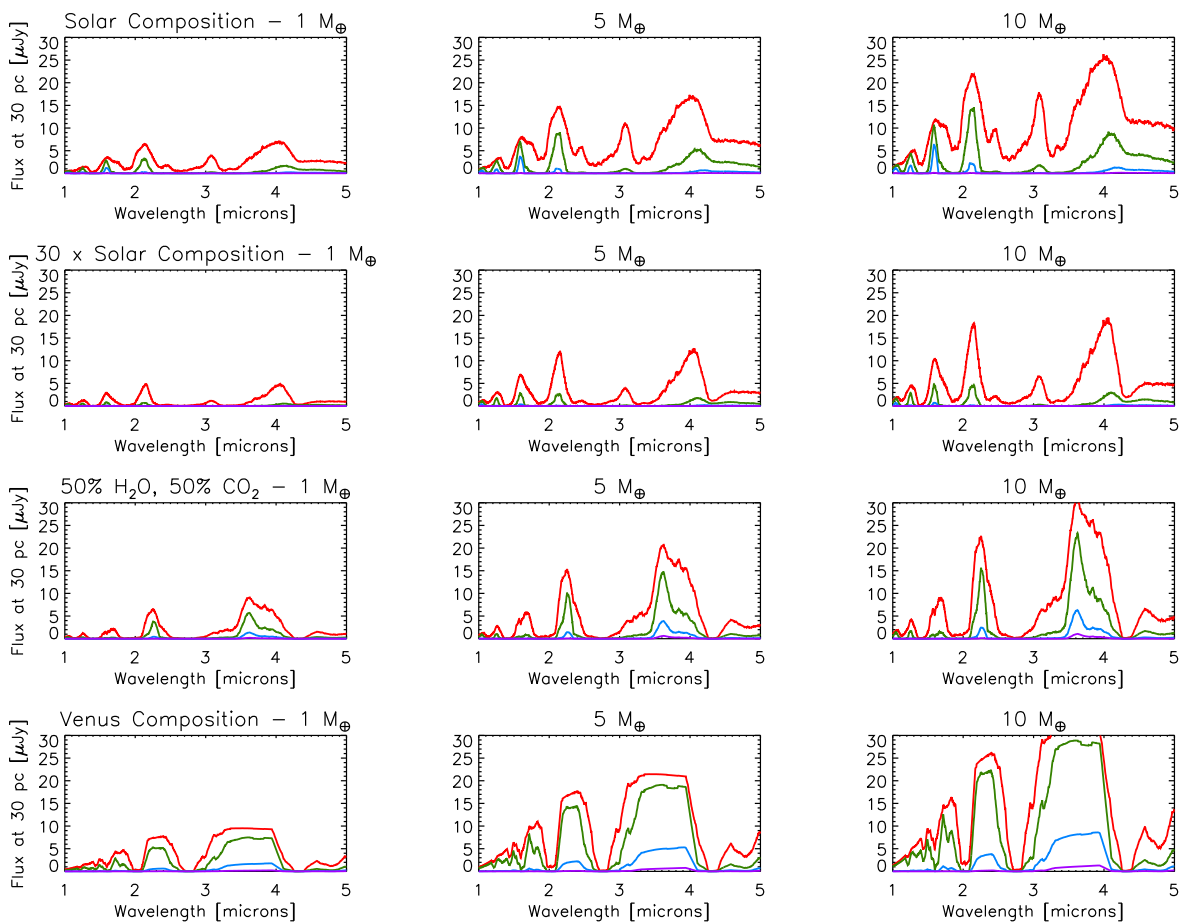}
\caption{Spectra of hot molten super-Earths for 4 of the 6 cases of atmospheric
        composition in flux units.  (Compositional scenarios 3 and 5 have been
	omitted here, as their spectra are quite similar to those of
	case 4 -- 50\% H$_2$O, 50\% CO$_2$.) The more massive planets give off 
	higher overall fluxes due to their larger surface areas.  
        \label{flux}}
\end{figure}
\begin{figure}
\plotone{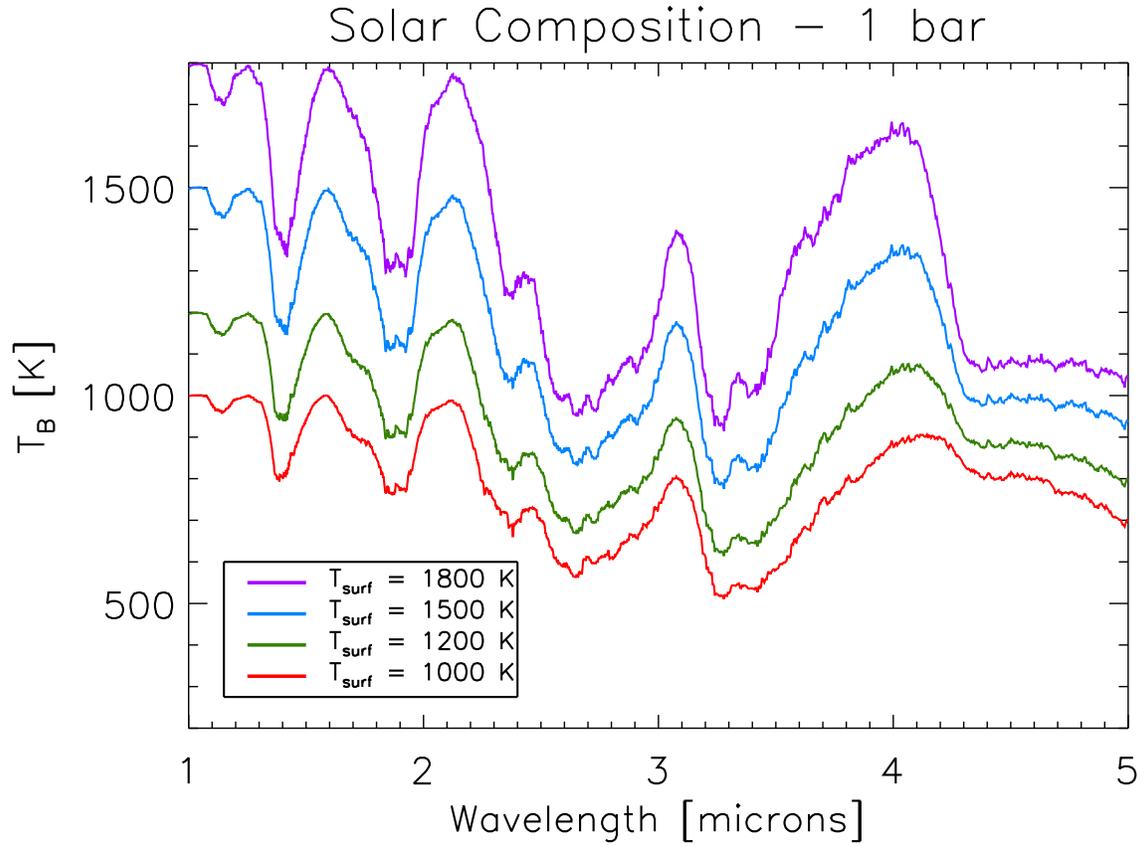}
\caption{Sample spectra of hot molten super-Earths showing the dependence
        of the emergent spectrum on the planet's surface temperature.  Spectra
	for planets with 1800, 1500, 1200, and 1000 K surface temperatures
	are plotted in purple, blue, green, and red, respectively, for a
	1-bar solar-composition atmosphere.  
        \label{temp}}
\end{figure}
\begin{figure}
\plotone{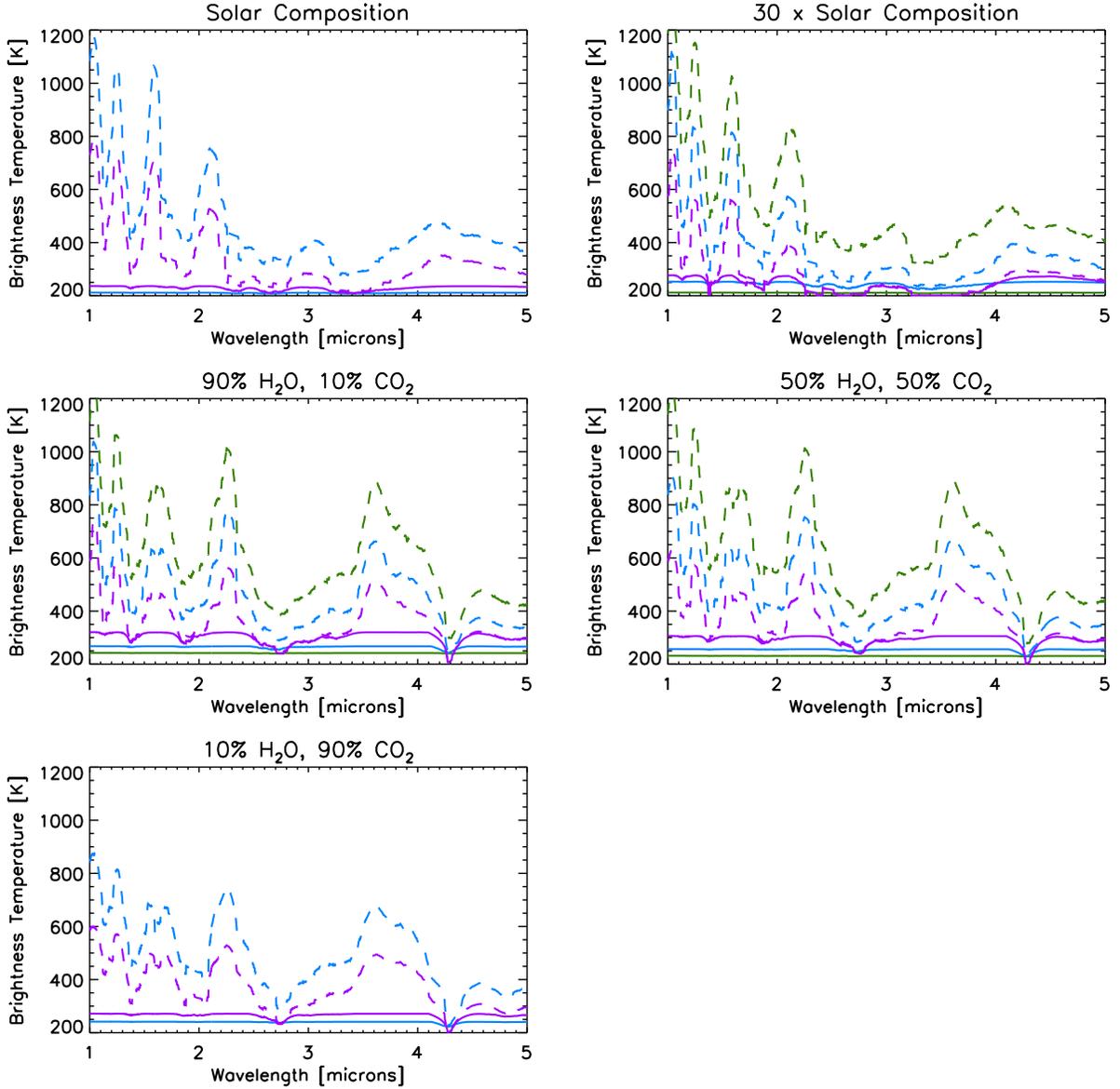}
\caption{Emission spectra for hot super-Earths ($T_{surf}$ = 1500 K) with 
        cloudy and partially cloudy atmospheres.  It is assumed that the clouds
	here are due to water condensation.  Solid-lined spectra denote the 
	case where the entire planet is blanketed by clouds, whereas the dashed
	lines indicate cases where the clouds only obscure 75\% of the 
	planetary surface.  Planets with surface pressures of 10, 100, and 1000
	bars are denoted by green, blue, and purple lines respectively.  None 
	of the 1-bar atmospheres are expected to have clouds owing to the fact 
	that their temperature-pressure profiles do not cross the water 
	condensation curve.  Additionally, none of the Venus-composition 
	atmospheres are expected to have water clouds for the same reason.  The
	maximum brightness temperature for an atmosphere with 
	thick water clouds is only 320 K.  However, if clouds are patchy 
	and obscure only a fraction of the planet, then the
	possibility exists to observe much higher brightness temperatures -  
	potentially as high as 1200 K if the clouds only cover 75\% of the
	planetary surface.  
        \label{clouds}}
\end{figure}

Once astronomers do succeed in imaging protoplanet collision afterglows, 
follow-up observations of these objects will additionally allow for
characterization of their atmospheres.  The bulk 
composition of super-Earth atmospheres is currently
an open question, owing to the fact that no observations currently exist
to constrain their properties.  Theoretical models provide some insight into
the expected range of atmospheric composition, but even these models
produce a broad range of possible atmospheres \citep[e.g.][]{elk08b, gen03}.
Indeed, young super-Earths are likely host to a wide variety of 
atmospheres, with bulk compositions that depend strongly on the planet's
formation history.  
Fortunately, many of the planet imagers planned for next-generation telescopes
are designed to have spectrographic capabilities -- necessary for 
characterizing  the atmospheres of these young protoplanets.  However, one look
at Figures~\ref{spectra}-\ref{clouds} should be enough to convince the reader
that there is significant degeneracy between the emergent spectra for the
various atmosphere scenarios that we have proposed.  Unambiguously determining
the atmospheric composition, mass fraction, and surface temperature for a 
collisional afterglow will therefore require spectral observations taken at a
high enough spectral resolution and signal-to-noise to break these 
degeneracies.  

One particularly interesting question that could be addressed by spectral 
observations of super-Earth collision afterglows is to determine whether or not
young silicate planets can form initially ``dry'' atmospheres (those without 
water).  In the study of solar system planets, there is ongoing debate as to 
whether or not Venus formed dry \citep[e.g.][]{lew72, gri88} or lost its 
hydrogen later in its lifetime due to photodissociation of water vapor and 
hydrodynamic escape \citep[e.g.][]{kas88}.  The atmosphere models presented
in this paper produce markedly different spectra for
a dry Venus-composition atmosphere and a wet atmosphere of 10\% water.  In 
particular, these spectra differ by factors of 10 or more in emitted flux
in the wavelength ranges of 1.1-1.2, 1.3-1.5, 1.8-1.9, 2.4-2.6, and 3.0-3.4 
$\mu m$.  Observations taken in any of these wavebands at a signal-to-noise of
several should therefore be sufficient to differentiate between wet and dry
atmospheres, potentially allowing astronomers to determine whether dry 
atmospheres on young super-Earths are a possibility.  

\acknowledgements 

We would like to thank Bruce Macintosh for sharing his insight on exoplanet 
imaging techniques and Dan Fabricant for discussions of the imaging 
capabilities of the GMT.  We also thank Phil Hinz, Scott Kenyon, and Eric
Mamajek for 
useful discussions.  MRM acknowledges the Harvard Origins of Life Initiative, 
the Smithsonian Astrophysical Observatory, and a NASA TPF Foundation Science 
Program grant NNG06GH25G (PI: S. Kenyon) for sabbatical support.  LE-T 
acknowledges funding from the NSF astronomy program.  

\bibliography{ms}



\end{document}